\documentclass[nofootinbib]{revtex4}
\usepackage{amsmath}
\usepackage{graphicx}
\newcommand{\be}{\begin{equation}}

\newcommand{\ee}{\end{equation}}

\newcommand{\beq}{\begin{eqnarray}}
\newcommand{\eeq}{\end{eqnarray}}

\begin{document}

\title{Dispersion relation of the nonlinear Klein-Gordon equation through a 
variational method }
\author{Paolo Amore}
\email{paolo@ucol.mx}
\affiliation{Facultad de Ciencias, Universidad de Colima, Bernal D\'{i}az del 
Castillo 340, Colima, Colima, M\'exico.}
\author{Alfredo Raya}
\email{raya@nucleares.unam.mx}
\affiliation{Facultad de Ciencias, Universidad de Colima,
Bernal D\'{i}az del Castillo 340, Colima, Colima, M\'exico}
\affiliation{Instituto de Ciencias Nucleares, Universidad Nacional Aut\'onoma
  de M\'exico. Circuito Exterior, C. U., A. Postal 70-543, M\'exico, D. F., 
04510}

\begin{abstract}
We derive approximate expressions for the dispersion relation of the
nonlinear Klein-Gordon equation in the case of strong nonlinearities using
a method based on the  Linear Delta Expansion. All the
results obtained in this article are fully analytical, never involve the use of
special functions, and can be used to obtain  systematic approximations to the
exact results to any desired degree of accuracy. We compare our findings
with similar results in the literature and show that our approach leads to
better and simpler results. 
\end{abstract}

\maketitle

{\bf The nonlinear Klein-Gordon equation describes a variety of physical
phenomena such as dislocations, ferroelectric and ferromagnetic domain walls,
DNA dynamics and Josephson junctions. The simple sinusoidal solutions to  the  
linear wave equation, which provide a dispersion relation independent of the
amplitude, are lost when nonlinear terms are considered. As a matter of fact
the exact solutions cannot be expressed in a simple form in terms of their
linear counterparts, although  they may still be oscillatory. Moreover the
dispersion relation obtained in this case turns out to depend upon the
amplitude.  The solution of the nonlinear wave equation poses an interesting
challenge,  expecially in the case of strong nonlinearities, where
perturbation theory by itself is not applicable: indeed in such cases the
perturbative series does not converge  and no sensible information can be
extracted directly from it. Here we present a variational method based on the
linear delta expansion  to find fully analytical approximate dispersion
relations for the nonlinear Klein-Gordon  and the Sine-Gordon equations for
weak and strong  nonlinearities. Our method can be easily generalized to other
cases and  provides a systematic way to achieve the desired  degree of
accuracy. The solutions obtained in this paper are fully analytical and  never
involve the use of special functions.
}

\section{\label{sec:Intro} Introduction}

In this article we study the problem of describing the propagation of traveling
waves obeying the nonlinear Klein-Gordon and the Sine-Gordon equations. Under
certain conditions the effect of the nonlinearity is to preserve the
oscillatory behavior of the solutions and, at the same time, modify the dispersion relation for the traveling waves which turns out to depend on the amplitude of oscillation. As a matter of fact we are considering conservative 
systems, for which the dynamics can be mapped to the nonlinear  oscillation of
a point mass in a one--dimensional potential. The main goal of this article is
to explore the effects of the nonlinearity on the solutions,  providing simple
and efficient approximations. Although for weak nonlinearities,  this task can be accomplished by applying perturbative methods (corresponding  to performing an expansion in a small parameter which governs the strength of the  nonlinearity itself), the situation is more complicated in presence of strong
nonlinearities. In such a regime perturbation theory cannot be applied, since
the perturbative series do not converge.  

Such a problem was studied in Ref.~\cite{Lim}, where nonperturbative formulas
for the dispersion relations of the traveling wave in the Klein-Gordon and
the Sine-Gordon equations were derived. The formulas obtained by Lim  \emph{et
  al.} provide an accurate approximation to the exact results even when the
nonlinearity is very strong.  

In this article we consider the same problems of Ref.~\cite{Lim} and apply to
them an approach which has been developed recently.~\cite{AA03,AL04,AA03b,AS,AASF}  Our approach is fully
nonperturbative in the sense that it does not correspond to a polynomial in
the nonlinear driving parameter and, when applied to a given order, allows
us to obtain analytical expressions  for the dispersion relations, which never
involve special functions, to any desired level of accuracy.  It is worth 
mentioning that in the case of weak nonlinearities, an expansion of the
nonperturbative results in powers of the nonlinear parameter is sufficient
to  recover the perturbative results.

Let us briefly describe the problem that we are interested in. We consider the
nonlinear Klein-Gordon equation  
\begin{equation}
u_{tt}-u_{xx}+V'(u)=0\;,\label{eq:nlKGe}
\end{equation}
where $V'(u)$ is a function of $u$, which we will assume to be odd, and the
prime is the  derivative with respect to $u$. To determine the periodic
traveling wave, we set 
\begin{equation}
u=u(\theta)\,,\qquad \theta=kx-\omega t\;.
\end{equation}
After substituting into Eq.(\ref{eq:nlKGe}) we find
\begin{equation}
\Omega^2 \ddot{u}+V'(u)=0\;,\label{eq:map}
\end{equation}
where $\Omega^2=(\omega^2-k^2)$ and $\dot u \equiv du/d\theta$. $u(\theta)$ is
periodic with period $2\pi$ and fulfills the boundary conditions 
\begin{equation}
u(0)=A\;,\qquad \dot u(0)=0\;,\label{eq:bc}
\end{equation}
with $A$ being the amplitude of the traveling wave. The solution of
Eq.(\ref{eq:map}) with the previous boundary conditions oscillates between $-A$
and $A$. By integrating Eq.(\ref{eq:map}) and taking into account
Eq.(\ref{eq:bc}) we obtain 
\begin{equation}
\frac{1}{2} \Omega^2 \dot{u}^2+V(u)=V(A)\;.
\end{equation}
Considering $\Omega^2>0$ we observe that 
\begin{equation}
\Omega = \frac{\pi}{\sqrt{2}\int_0^A[V(A)-V(u)]^{-1/2}du}\label{eq:exact}
\end{equation}
gives the exact expression for the dispersion relation of the nonlinear
Klein-Gordon equation. We neglect the case $\Omega^2\le 0$ since there is no
traveling wave for this configuration. 

This article is organized as follows: in section \ref{sec_2} we describe the
variational nonperturbative  approach and apply it to derive approximate
analytical formulas for the  nonlinear Klein--Gordon equation; in section
\ref{sec_3} we apply our method to two further  nonlinear equations;
finally in section \ref{sec_4} we draw our conclusions.

\section{Variational Method}
\label{sec_2}

An exact solution of Eq.~(\ref{eq:nlKGe}) can be accomplished in  a limited
number of cases, depending on the form of the potential $V(u)$. However, when
the nonlinearities due to the potential $V(u)$ are small, it is still possible
to find useful approximations using perturbation theory. The focus of this
section will be on the opposite situation, when the nonlinearities are not
small and a perturbative expansion is not useful. In such a case one needs to
resort to nonperturbative methods, capable of providing the solution even 
in the presence of strong nonlinearities. One of such methods, which we
will use in the present article, is the linear delta expansion
(LDE)~\cite{K81,F00,AFC90,lde}. 

The LDE is a powerful technique that
has been applied to difficult problems arising in different branches of
physics  like field theory, classical, quantum and statistical mechanics.
The idea behind the LDE is to interpolate a given problem ${\cal P}_g$ with a solvable one ${\cal P}_s$, which depends on one or more arbitrary 
parameters $\lambda$. In symbolic form ${\cal P}={\cal P}_s(\lambda) +
\delta ({\cal P}_g-{\cal P}_s(\lambda))$. $\delta$ is just a bookkeeping
parameter such that for $\delta=1$ we recover the original problem, and for $\delta\to 0$ we can perform a perturbative expansion of the solutions of ${\cal P}$ in $\delta$. The perturbative solution obtained in this way to a finite order shows an artificial dependence upon the arbitrary parameter, $\lambda$, and would cancel if the calculation were carried out to all orders. As such we must regard  such dependence as unnatural; in order to minimize the spurious effects of $\lambda$ we then require that any observable  ${\cal O}$,
calculated to a  finite order, be locally independent on $\lambda$, i.e. that  
\begin{equation}
\frac{\partial {\cal O}}{\partial \lambda}=0\;.\label{eq:PMS} 
\end{equation}
This condition is known as  the ``Principle of Minimal Sensitivity''
  (PMS)~\cite{S81}. We call $\lambda_{PMS}$ the solution to this equation.
  (In the case where the PMS equation has multiple solutions, the  solution with  smallest second derivative is chosen.)  We emphasize that
  the results that we obtain by applying this method do not correspond to a
  polynomial in the  parameters of the model
as in the case of perturbative methods.

The procedure that we have illustrated is quite general and it will be
possible to implement it in different ways depending on the problem that is
being considered. In  Refs.~\cite{AA03,AL04,AA03b} the LDE was used in
conjunction with the Lindstedt--Poincar\'e technique to solve the
corresponding equations of motion. Our approach here is to apply 
the LDE directly to the integral  of eq.~(\ref{eq:exact}) as in
Refs.~\cite{AS} and~\cite{AASF}.  We will consider the potential 
\begin{equation}
V(u)=\frac{u^2}{2}+\frac{\mu u^4}{4}\label{eq:pot} \ .
\end{equation}

The dispersion relation in this case can be obtained using
Eq. (\ref{eq:exact}) as 
\begin{equation}
\Omega=\frac{\pi\sqrt{1-\mu A^2}}{2\int_0^\pi
  (1-m\sin^2{\phi})^{-1/2}d\phi}\;,
\end{equation}
with $m=\frac{\mu A^2}{2(1+\mu A^2)}$.

We consider the following  approach to obtain the dispersion relation of a
periodic traveling wave. This comes from the equation for the period of
oscillations, 
\begin{equation}
T = \int_{-A}^{+A} \frac{\sqrt{2}}{\sqrt{E-V(u)}}du\;,\label{eq:period}
\end{equation}
where the total energy $E$ is conserved and  $\pm A$ are the classical turning
points. 

In the spirit of the LDE we interpolate the nonlinear potential $V(u)$ with a
solvable potential $V_0(u)$ and define the interpolated potential
$V_\delta(u)=V_0(u)+\delta(V(u)-V_0(u))$. Notice that for $\delta=1$,
$V_\delta(u)=V(u)$ is just the original potential, whereas for $\delta=0$  it
reduces to $V_0(u)$. Hence we can write Eq. (\ref{eq:period})
as~\cite{AS,AASF} 
\begin{equation}
T_\delta = \int_{-A}^{+A} \frac{\sqrt{2}}{\sqrt{E_0-V_0(u)}} \frac{du}
{\sqrt{1+\delta \Delta(u)}}
\end{equation}
where
\begin{equation}
\Delta(u)=\frac{E-E_0-V(u)+V_0(u)}{E_0-V_0(u)}\;.
\end{equation}
Obviously $E=V(A)$ and $E_0=V_0(A)$.

We treat the term proportional to $\delta$ as a perturbation and expand
in powers of $\delta$. This allows us to write 
\begin{equation}
T_\delta = \sum_{n=0}^\infty \frac{(2n-1)!!}{n! 2^n}(-1)^n \delta^n
\int_{-A}^{+A}  
\frac{\sqrt{2} (\Delta(u))^n}{\sqrt{E_0-V_0(u)}}du\;. \label{eq:periodLDE}
\end{equation}

Observe that the integrals in each order of Eq.~(\ref{eq:periodLDE}) have
integrable singularities at the turning points because $\Delta(\pm A)$ is
finite. Assume that $|\Delta(u)|\leq \Delta_0 < 1$ for $u\in [-A,A]$, which happens if $\lambda$, the arbitrary variational parameter, is chosen appropiately. Then, the series (\ref{eq:periodLDE}) converges uniformly for $|\delta | < 1/\Delta_0$, which includes the case $\delta=1$.

For the potential given in Eq. (\ref{eq:pot}) we can choose
$V_0(u)=\frac{1+\lambda^2}{2}u^2$ as the interpolating potential and hence we
have 
\begin{equation}
\Delta(u)=\frac{2}{1+\lambda^2}\left[ \frac{\mu}{4}(a^2+u^2)
-\frac{\lambda^2}{2}\right]\;.
\end{equation}
The parameter $\lambda$ should be chosen to be  $\lambda>\sqrt{\frac{\mu
    A^2}{2}}\sqrt{1+\frac{1}{\mu A^2}}$ which 
guarantees the uniform convergence of Eq.(\ref{eq:periodLDE}).

It is straightforward to check that at first order,
\begin{equation}
T_\delta^{(0)}+\delta T_\delta^{(1)}= \frac{2\pi}{\sqrt{1+\lambda^2}}
\left\{1-\frac{\delta}{1+\lambda^2}\left[ \frac{3}{8} 
\mu A^2 -\frac{\lambda^2}{2}\right] \right\}\;.
\end{equation}

The PMS (\ref{eq:PMS}) with ${\cal O}=T$ yields
\begin{equation}
\lambda_{PMS}=\frac{\sqrt{3\mu}A}{2}\;.\label{eq:lpms_t}
\end{equation}
The period is found to be  
\begin{equation}
T_{PMS}=\frac{4\pi}{\sqrt{4+3\mu A^2}}\;.
\end{equation}
Correspondingly,
\begin{equation}
\Omega_{LDE\,(1)}=\sqrt{1+\frac{3}{4} \mu A^2}\;.\label{eq:omega_period}
\end{equation}

In Ref.~\cite{AASF} it was found that with this value of $\lambda_{PMS}$ all
the remaining terms of odd order in Eq.~(\ref{eq:periodLDE}) vanish. Hence,
retaining only nonvanishing contributions, the expression for the period at
order $N$ is 
\begin{equation}
T_\delta^{(N)}= 
\frac{4\pi}{\sqrt{4+3 \mu A^2}} \sum_{n=0}^N (-1)^n \left( \begin{array}{c} 
-1/2\\n\end{array}\right)
\left( \begin{array}{c} -1/2\\2n\end{array}\right) \left(\frac{\mu A^2}
{4+3\mu A^2} \right)^{2n}\;,
 \label{eq:T_ordn}
\end{equation}
and, correspondingly,
\begin{equation}
\Omega_{LDE\,(N)}=\frac{2\pi}{T_\delta^{(N)}}\;.  \label{eq:W_ordn}
\end{equation}
At second order we have
\begin{equation}
\Omega_{LDE\,(2)}=\frac{\sqrt{4+3 A^2\mu}}{2 \displaystyle{\left(1+\frac{3
     A^4 \mu^2(1024 + 
     A^2\mu(1536+611  A^2\mu))}{1024 (4+ 3 A^2\mu)^4} \right)}}\;.
\label{eq:omegaper_2}
\end{equation}
At third order, the dispersion relation is given by
\begin{equation}
\Omega_{LDE\,(3)}= \frac{{\sqrt{4 + 3\,A^2\,\mu }}}
  {2\,\displaystyle{\left( 1 + \frac{3\,A^4\,{\mu }^2\,
         \left( 385\,A^8\,{\mu }^4 + 
           560\,A^4\,{\mu }^2\,
            {\left( 4 + 3\,A^2\,\mu  \right) }^2 + 
           1024\,{\left( 4 + 3\,A^2\,\mu  \right) }^4 \right) }{
         16384\,{\left( 4 + 3\,A^2\,\mu  \right) }^6} \right)} }\;.
\label{eq:omegaper_3}
\end{equation}

We will compare the results obtained using our method,
Eqs. (\ref{eq:omega_period}), (\ref{eq:omegaper_2}), and (\ref{eq:omegaper_3})
with the results obtained  in Ref. \cite{Lim}, where the same problem has been solved using the harmonic balance technique in combination with the linearization of the nonlinear Klein-Gordon equation. The findings of Ref. \cite{Lim} at first order, their expression for the dispersion relation coincides with our Eq.(\ref{eq:omega_period}), whereas at the second order they find
\begin{equation}
\Omega_{Lim\,(2)} = \sqrt{\frac{40+31\mu A^2 + \sqrt{1024+1472\mu A^2 
+ 421 \mu^2 A^4}}{72}}\label{eq:Lim2}.
\end{equation}

\begin{figure}%[tbp]
\begin{center}
\includegraphics[width=14cm]{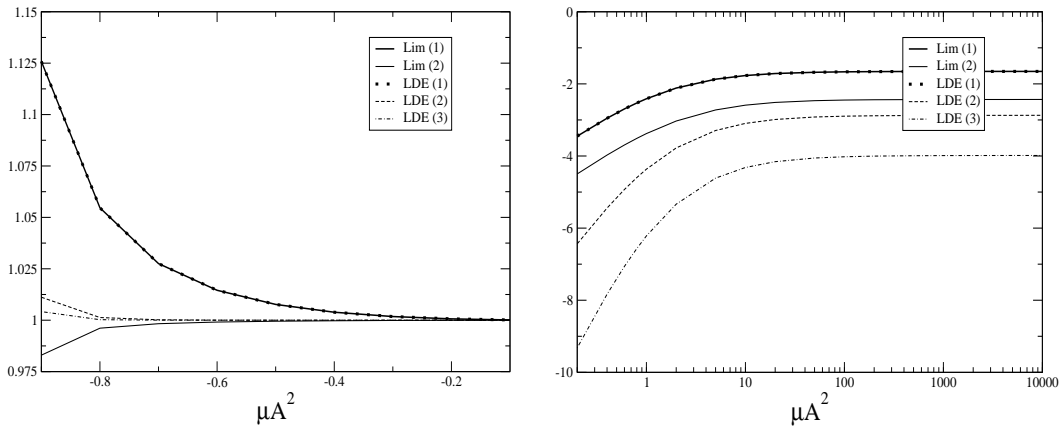}
\end{center}
\caption{(Left) Ratio of the dispersion relation from Eqs.~
(\ref{eq:omega_period}) and (\ref{eq:omegaper_2})-(\ref{eq:Lim2}) exactly for $\mu A^2<0$ and (right) 
relative error $\Delta$ [see eq.~(\ref{relerr})] of the dispersion relations
for $\mu A^2\gg 0$.} 
\label{fig1}
\end{figure}

In the left-hand panel of Fig.~\ref{fig1} we make a comparison of the ratios of the dispersion relations obtained from Eqs.~(\ref{eq:omega_period}), and
(\ref{eq:omegaper_2})-(\ref{eq:Lim2})  to the
exact dispersion relations for $\mu A^2<0$, and in the right-hand panel of
Fig.~\ref{fig1} we display the relative error 
\begin{equation}
\Delta=\log_{10}\left|\frac{
    \Omega-\Omega_{exact} }{\Omega_{exact}}\right|\label{relerr}
\end{equation} 
for $\mu A^2\gg 0$. We can appreciate that our variational method at second
order provides a  smaller error than the method of  Ref. \cite{Lim} applied to
the same order.  The error is further reduced by using the LDE to the third
order and can be then  systematically reduced using the general formula
(\ref{eq:T_ordn}).  

\section{Further Examples}
\label{sec_3}

\subsection{Sine-Gordon model}

We now consider the Sine-Gordon model, which is governed by the potential
\begin{equation}
V(u)=-\cos{u}\label{eq:sGpot}
\end{equation}
and which allows us to write the nonlinear Klein-Gordon equation, also known
as the Sine-Gordon equation as 
\begin{equation}
\Omega^2 \ddot u+ \sin{u}=0\;.\label{eq:sGeq}
\end{equation}
The exact dispersion relation in this case can be obtained from
\begin{equation}
\Omega = \frac{\pi}{2\int_0^{\pi/2}(1-m^2 \sin^2{t})^{-1/2}dt} 
\label{eq:oexact_sG}
\end{equation}
with $m=\sin(A/2)$. Observe that in this case 
\begin{equation}
T=4 \int_0^{\pi/2}(1-m^2 \sin^2{t})^{-1/2}dt \equiv 4K(m^2)\;,
\end{equation}
with $K(m)$ being the elliptic integral of the first kind. We take advantage of
this fact and make use of the nonperturbative series for the elliptic
integral which was derived using the LDE technique~\cite{chavos}. At order
$N$, setting $\lambda=-m/2$ and $\delta=1$, it is given by the expression: 
\begin{equation}
K_N(m,\lambda)= \frac{\pi}{2} \ \sum_{k=0}^N \sum_{j=0}^k
\frac{\Gamma(j+1/2)}{j!^2 \,(k-j)!\,  
\Gamma(1/2-k)} \ \frac{(-m)^k }
{2^{k-j} \ (1-\frac{m}{2})^{k+1/2}} \;.
\end{equation}
This expression provides a nonperturbative series for the elliptic integral
of the first kind since it does not correspond to a simple polynomial in $m$.

To further improve this series we can use the Landen transformation
\cite{AbrSte}
\begin{equation}
K(m)=\frac{1}{1+\sqrt{m}} K\left( \frac{4\sqrt{m}}{(1+\sqrt{m})^2}\right)
\label{eq:landel}
\end{equation}
and the inverse relation
\begin{equation}
K(m)=\frac{2(1-\sqrt{1-m})}{m} K \left( \frac{(-2+2\sqrt{1-m}+m)^2}
{m^2}\right)\;.\label{eq:landeninv}
\end{equation}
Notice that $f(m)= \frac{4\sqrt{m}}{(1+\sqrt{m})^2}$ maps a value $0<m<1$ into
a new value $m'=f(m)>m$. The inverse transformation $f^{-1}(m)=
\frac{(-2+2\sqrt{1-m}+m)^2}{m^2}$ maps a value $m$ into a smaller one. Using
this transformation we obtain more accurate approximations for the elliptic
integrals. For example, at order 1 we find   
\begin{equation}
K_{LDE\,(1)}(m) = \frac{\pi}{\sqrt{1-\frac{m}{2}+3\sqrt{1-m}}}
\end{equation}
and, correspondingly
\beq
\Omega_{LDE\,(1)} &=&  \frac{1}{4} \sqrt{\cos (A)+12
  \left|\cos\left(\frac{A}{2}\right)\right|+3} 
\;.\label{eq:oLDE}
\eeq
At second order we find
\begin{eqnarray}
\Omega_{LDE\,(2)} &=& 
\frac{16 \cos^2{\left(\frac{A}{4} \right)} (3+12\cos{\left(\frac{A}{2} \right)
  + \cos(A)})^2 \sqrt{2+2\cos{\left(\frac{A}{2}
    \right)}\sec^4{\left(\frac{A}{4} \right)} } }{2713+2520
\cos{\left(\frac{A}{2}\right) + 2580 \cos{(A)} + 360 \cos{\left( \frac{3A}{2}
\right)      + 19 \cos(2A)} }}
\;\label{eq:sgomega_3}
\end{eqnarray}
It is noticeable that $\Omega_{LDE\,(3)}=\Omega_{LDE\,(2)}$. In fact, for the
following consecutive orders, the same statement holds, i. e.,
$\Omega_{LDE\,(5)}=\Omega_{LDE\,(4)},\, \Omega_{LDE\,(7)}=\Omega_{LDE\,(6)}$
and so on. The same pattern of equal value of the observables for consecutive
orders of approximation was found in Ref.~\cite{AL04} for the Duffing
potential at large $n. $

For comparison, Lim \emph{et al.}~\cite{Lim} have found the dispersion
relation to be given at first order as 
\begin{equation}
\Omega_{Lim (1)} = \sqrt{\frac{2J_1(A)}{A}}\;,\label{sGLim1}
\end{equation}
and at second order as
\begin{equation}
\Omega_{Lim (2)}=\sqrt{g(A)+\sqrt{g^2(A)-h(A)}}\;,\label{sGLim2}
\end{equation}
where
\begin{eqnarray}
g(A)&=& \frac{(b_0-b_2-b_4+b_6)A+18a_1+2a_3}{36 A}\nonumber\\
h(A)&=& \frac{a_1(b_0-b_2-b_4+b_6)}{18A}
\end{eqnarray}
and
\begin{equation}
a_1=2J_1(A),\quad a_3=-2J_3(A),\quad b_{2i}=2(-1)^i J_{2i}(A),\quad i=0,1,2,3
\end{equation}
$J_n(A)$ being the Bessel function of the first kind.

\begin{figure}%[tbp]
\begin{center}
\includegraphics[width=14cm]{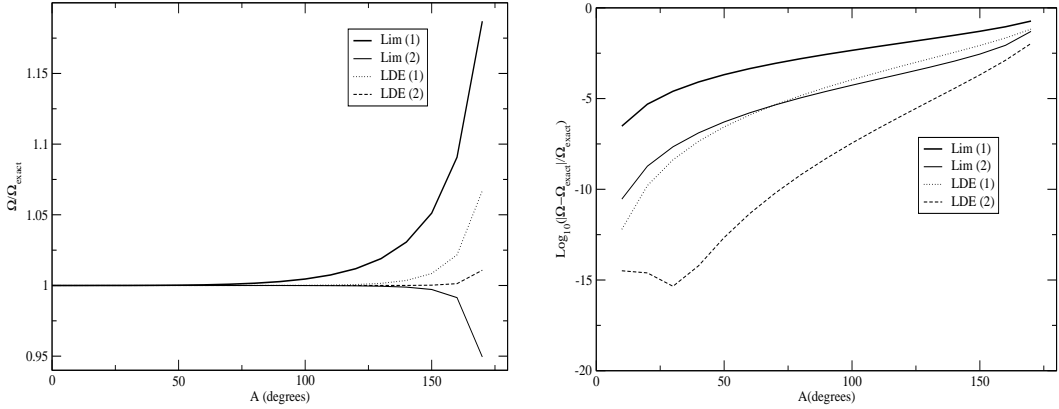}
\end{center}
\caption{(Left) Ratio of the dispersion relation from eqs.~
(\ref{eq:oLDE})-(\ref{sGLim2}) exactly and (right)  
relative error $\Delta$ [see eq.~(\ref{relerr})] of the dispersion relations.} 
\label{fig2}
\end{figure}

In the left-hand panel of Fig.~\ref{fig2} we display the ratio of the dispersion
relations from  Eqs.~(\ref{eq:oLDE})-(\ref{sGLim2}) to the exact and on the right-hand panel the corresponding relative
errors. From the graphs we see that the LDE curves  calculated to second order
display much smaller errors than the curves obtained with the method of Lim
\emph{et al.} even close to  $A= \pi$. A second observation is that our
formulas can be systematically  improved simply by going to a higher order and
that they do not involve any special function, as in the case of
Eq.~(\ref{sGLim1}).

\subsection{Pure quartic potential}

Our final example is the Klein-Gordon equation in a pure quartic potential 
\begin{equation}
V(u)=\frac{u^4}{4}\;,
\end{equation}
which leads to the equation of motion
\begin{equation}
\ddot u + u^3=0\;.\label{eq:pure_cub}
\end{equation}

This is a particular case of the first example where the contribution of the
quadratic term in the potential~(\ref{eq:pot}) is neglected. As such, the
corresponding dispersion relation can be derived from the expression of the
period of oscillations, Eq.~(\ref{eq:T_ordn}), since the quadratic term
contributes with the $4$ in the square root in the front of the double sum and
in the argument in the sum, and is simply given by 
\begin{equation}
T_{LDE\,(N)} = \frac{4\pi}{\sqrt{3 \mu A^2}}\sum_{n=0}^N (-1)^n \left( 
\begin{array}{c} -1/2 \\ n \end{array}\right)  \left( \begin{array}{c} -1/2 
\\ 2n \end{array}\right) \frac{1}{3^{2n}}\; 
\end{equation}
and correspondingly $\Omega_{LDE\,(N)}$ can be obtained as in
Eq.~(\ref{eq:W_ordn}).

Results for the first three orders are the following~:
\begin{eqnarray}
\Omega_{LDE\,(1)}&=& \frac{24 \sqrt{3}A}{49}\;,\label{eq:cubOrd1}\\
\Omega_{LDE\,(2)}&=& \frac{13824 \sqrt{3}A}{28259}\;,\label{eq:cubOrd2}\\
\Omega_{LDE\,(3)}&=& \frac{1990656 \sqrt{3}A}{4069681}\;. \label{eq:cubOrd3}
\end{eqnarray}

Lim \emph{et al.} found, at first and second order of approximation,
respectively, 
\begin{eqnarray}
\Omega_{Lim (1)}&=& \frac{\sqrt{3}}{2}A\;,\nonumber\\
\Omega_{Lim (2)}&=& \frac{1}{12}\sqrt{62+2\sqrt{421}}A\;.\label{eq:LIM}
\end{eqnarray}

\begin{figure}%[tbp]
\begin{center}
\includegraphics[width=14cm]{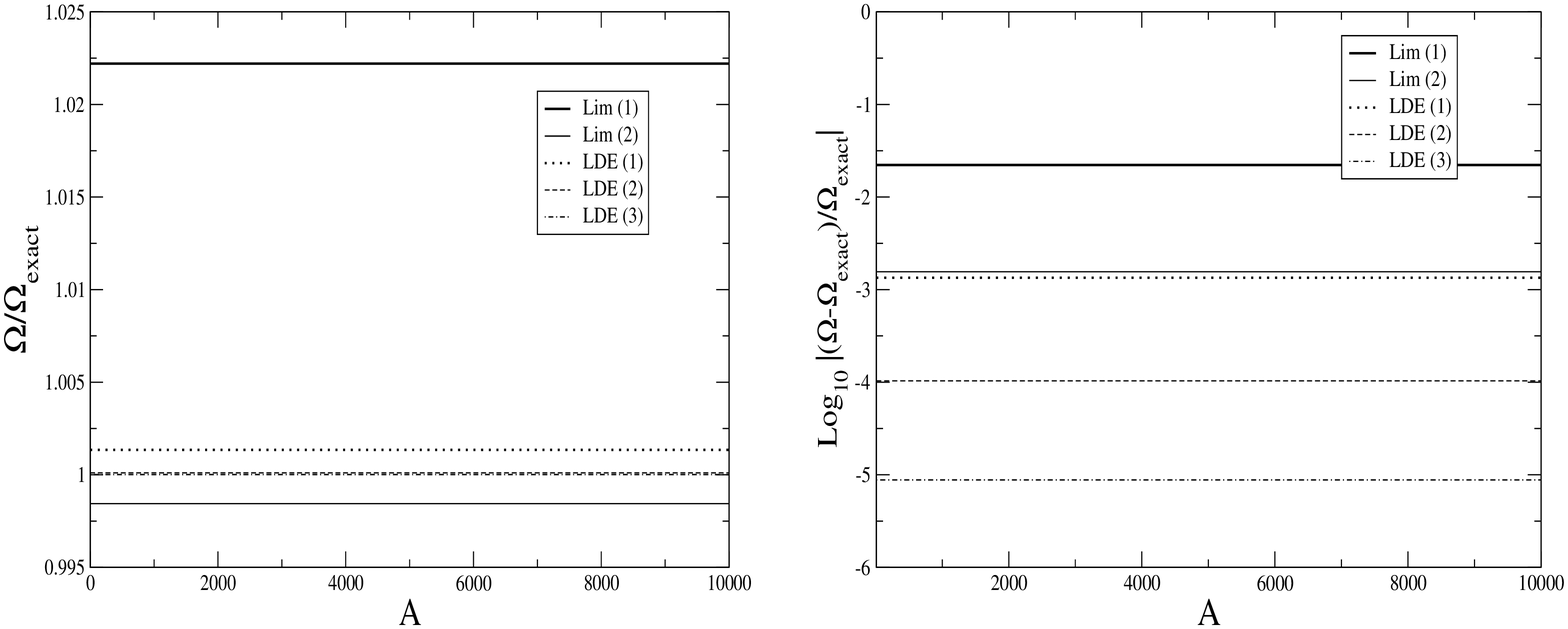}
\end{center}
\caption{(Left) Ratio of the dispersion relation from Eqs.~
(\ref{eq:cubOrd1})-(\ref{eq:cubOrd3}), and both
expressions in Eq.~(\ref{eq:LIM}) exactly and (right)  
relative error $\Delta$ (see eq.~(\ref{relerr})) of the dispersion  relations.}
\label{fig3}
\end{figure}

In the left-hand panel of Fig.~\ref{fig3}  we display the ratio of the approximate  to the exact dispersion relation and in the right-hand panel the relative error  from our  findings at first, second and third order and those of
 Ref.~\cite{Lim} given   previously.  At first order, our findings perform just as the second order of Lim  \emph{et al.}~\cite{Lim}, and at second and third orders, the performance of the  variational results is excellent.

\section{Conclusions}
\label{sec_4}

We have derived analytical expressions for the dispersion relations of the
nonlinear Klein-Gordon equation for different potentials by means of the
Linear Delta Expansion. This technique is implemented  by computing the period
of oscillations in the given potential. In the particular example of the
Sine-Gordon potential, where the dispersion relation is given in terms of
elliptic integrals, we have implemented the LDE to compute such integral and,
by means of the Landen transformation, we have obtained an improved series
for the elliptic integral. We have observed that the expression obtained by
using  the first few terms in this series performs remarkably well even close
to the $A = \pi$. We believe that our results are appealing in two  respects:
first in that they provide a systematic way to approximate the  exact result
with the desired accuracy, and second in that the expressions  that we obtain
never involve special functions, as in the case of  Ref.~\cite{Lim}. An aspect
that needs to be underlined is that the method  described in subsection
\ref{sec_2} provides a {\sl convergent} series  representation for the
dispersion relation, provided that the arbitrary  parameter fulfills a simple
condition.

\begin{acknowledgments}
One ofthe authors (P.A.) acknowledges the support of Conacyt grant no. C01-40633/A-1. The authors also acknowledge support  of the Fondo Ram\'on Alvarez Buylla of Colima University.
\end{acknowledgments}

\end{document}